# Self-assembled line network in BiFeO$_3$ thin films

B. Colson[a], V. Fuentes [a], Z. Konstantinović[b], D. Colson[c], A. Forget[c], N. Lazarević[b], M. Šćepanović[b], Z. V. Popović[b], C. Frontera [a], Ll. Balcells [a], B. Martinez [a], A. Pomar[a]

[a] Institut de Ciencia de Materials de Barcelona, ICMAB-CSIC, Campus UAB, 08193 Bellaterra, Spain
[b] Center for Solid State Physics and New Materials, Institute of Physics Belgrade, Pregrevica 118, University of Belgrade, Serbia
[c] SPEC, CEA, CNRS UMR 3680, Université Paris-Saclay, 91191 Gif sur Yvette Cedex, France

ABSTRACT

In this work we report on the controlled fabrication of a self-assembled line network in highly epitaxial BiFeO$_3$ thin films on top of LaAlO$_3$ in the kinetically limited grown region by RF sputtering. As previously shown in the case of manganite thin films, the remarkable degree of ordering is achieved using vicinal substrates with well-defined step-terrace morphology. Nanostructured BiFeO$_3$ thin films show mixed-phase morphology. Besides typical formation following (100) and (010) axes, some mixed phase nanodomains are detected also in-between the regular line network. These particular microstructures open a playground for future applications in multiferroic nanomaterials.

1. Introduction

Bismuth ferrite BiFeO$_3$ (BFO) is a very active research domain due to environment friendly room-temperature multiferroic character with wide range of potential applications, from the low-power spintronic to optical devices [1, 2]. The very large electrical polarization [3, 4], coupling between the polarization and magnetic easy plane [5, 6, 7] and its strong sensitivity on the epitaxial strain are crucial parameters for applications and for understanding their fundamental properties in general [8, 9]. The control over the ferroelectric polarization through the structural strain and the miscut angle of underlying substrates rapidly gain interest [10], additionally enhanced with the discovery of mixed phase nanodomains [9, 11, 12, 13]. In addition, it has been recently shown that the kinetic growth conditions allow synthesizing high quality films with selective ferroelectric domains [14].

Self-organization of long-range ordered nanostructures of transition metal oxide thin films is of major relevance for both, the study of enhanced or novel physical properties at the nanoscale (from enhanced magneto-resistive properties to unexpected interfacial effects) and for developing a new generation of devices [15, 16]. This bottom-up nanostructural approach presents an alternative to the more conventional top-down lithography-based methods with numerous advantages ranging from rapid preparation of low-cost and large surface oxide nanotemplates to the formation of nanoobjects with size and densities beyond actual possibilities [17]. Finally, ferroelectric BFO thin films crystallize in the very similar rhombohedral structure as previously studied LSMO thin films [15, 17] and open a huge playground for the formation of nanostructured networks at the surface.

In this paper, we report on the formation of regular nanostructures in ferroelectric BFO thin films grown on top of LaAlO$_3$ (LAO) substrates by RF sputtering. The nanostructuration of BFO thin films is directly induced by structural and morphological features of the underlying substrate (lattice parameter inducing strain conditions on the one side and step-terrace morphology and chemical affinity on the other side). In addition to regular network, stripe-like features are detected and analyzed by X-ray, Raman and AFM spectroscopy.

2. Experimental

BFO thin films were grown by RF magnetron sputtering on top of LaAlO$_3$ (001) substrates under an oxygen partial pressure of 0.19 Pa using commercial stoichiometric target (Kurt J. Lesker Company). Growth conditions (growth rate of F~0.03 ML/s and high growth temperature) were adjusted to promote self-organized surface nanostructures in the kinetic growth regime, i.e. far away from thermodynamic relaxation mechanisms, that have been previously studied in detail in GeSi semiconductors [18]. By a fine tuning of the growth kinetic pathway, the surface diffusion was reduced (but not completely suppressed), taking advantage of the unusual misfit strain relaxation in presence of stepped substrate [17]. Substrates were previously washed in milliQ water and thermally treated at 1000 °C to assure the presence of terrace-step morphology with unit cell height. The thickness value, $t$, of the different BFO/LAO films presented in this study is in the range of 40 nm <$t$<50 nm.

The surface morphology of the films was studied using Atomic Force Microscopy (AFM) and Scanning Electron Microscopy (SEM). AFM images were obtained in a MFP3D Asylum AFM while SEM images were obtained with a QUANTA FEI 200 FEG-ESEM. The crystal structure was characterized by X-ray diffraction (XRD) and reflectivity techniques (XRR) using a Siemens D5000 diffractometers with Kα-Cu radiation.

Magnetic characterization was performed at room temperature (in-plane configuration H∥(100) and out-of-plane configuration H⊥(100)) with a superconducting quantum interference device magnetometer (Quantum Design). In order to estimate the magnetization of the film, the diamagnetic background of substrates was subtracted (estimated from negative slope of M(H) at high magnetic field, 10000<H<50000 Oe).

Raman scattering measurements were performed using a Jobin Yvon T64000 Raman system in µ-Raman configuration. A Coherent VerdiG solid state laser with 532 nm line was used as an excitation source. Laser beam focusing was accomplished by a microscope objective with ×50 magnification.

3. Results and discussion

BiFeO$_3$ presents, at room temperature, a rhombohedral structure in bulk form (lattice constant $a_{bulk}$=3.964 Å) [3, 5]. In thin films, structural and functional properties of this multiferroic compound can be drastically modified due to presence of the structural strain induced by the selected substrate [19]. Particularly, the epitaxial growth of BFO on top of LAO substrate induces huge compressive in-plane strain, which allows stabilizing the tetragonal phase, correlated with theoretically predicted giant ferroelectric polarization [20]. In this study, the films were grown under large compressive strain of $\varepsilon=(a_{LAO}-a_{BFO})/a_{BFO} = -4.62$ %, giving rise to a significantly larger perpendicular cell parameter compared to bulk counterpart (see below).

Fig. 1 shows the formation of the long range ordered line network on top of the BFO surface during crystal growth. Fast Fourier Transform (FFT) (Fig.1(b)) of topographic AFM image (Fig.1(a)) indicate long-range order of grooves along one specific direction (see two high intensity dots in circles) with a separation around $1/k_\perp$~150 nm. Two additional dots (see arrows) are also visible (doubled distance in $k$ space) in possible correlation with the formation of the polarization nano-domains (see below). The regular lines, i.e. grooves have typical depth of around 6±2 nm (Fig.1(c)).

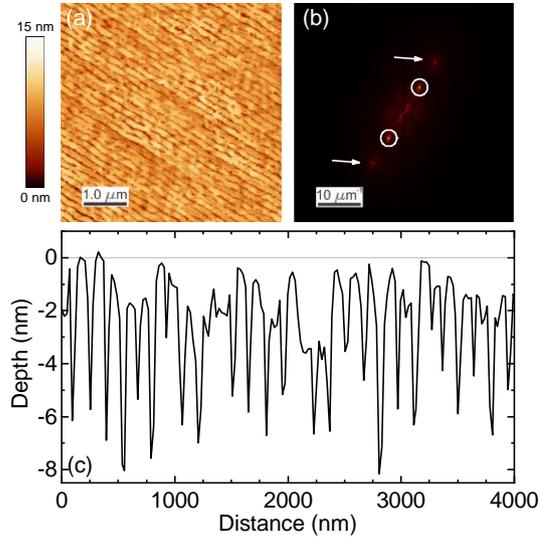

Fig. 1. (a) Topographic AFM image (4x4 μm$^2$) of BiFeO$_3$ film (thickness, $t$=45 nm) grown on top of LAO. and (b) corresponding Fast Fourier Transform (c) Typical line profile of regular grooves.

In order to further clarify the structural arrangement of the regular lines the topography of BFO thin films (Fig.2(b)) is compared directly with the topography of the underlying LAO substrate (Fig.2(a)). The corresponding height-height autocorrelation functions are shown in Fig.2(c) for LAO and (d) for BFO topography. The regular line pattern is visible at the overall surface in both cases, also as non-vanishing oscillations in the corresponding profile perpendicular to lines. A clear correlation between the two patterns is evident from the corresponding profile lines in Fig.2(e), demonstrating a typical separation between ordered nanostructured lines of l~140 nm, in agreement with the underlying terrace-step morphology of the LAO substrate with a miscut angle of α~0.155°.

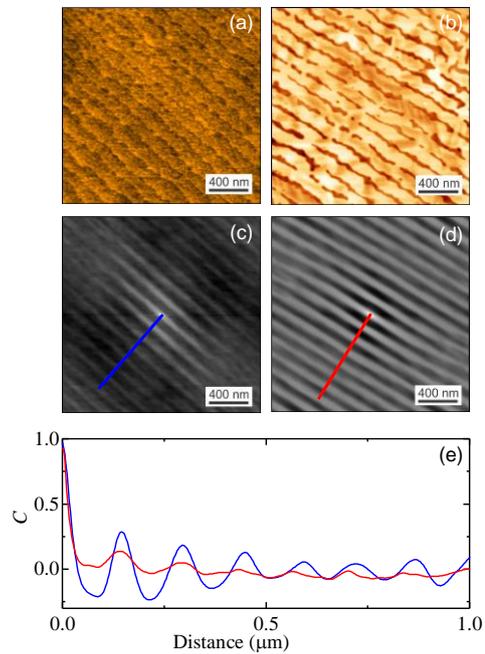

Fig. 2. (a) Topographic AFM images (2x2 μm$^2$) of (a) LaAlO$_3$ substrates (α~0.155°) and (b) BiFeO$_3$ film (thickness, $t$=45 nm) grown on top of it. The corresponding height-height correlation functions of the topographic surfaces are shown in (c) and (d) respectively. (e) Corresponding profile lines of LAO (blue) and of BFO (red).

The structural properties have been investigated by θ-2θ X-ray diffraction (XRD), Fig.3(a) shows scans around LAO (0 0 2) reflection. The strongest peak in Fig 3(a) corresponds to the LAO substrate (2θ =48.05° with lattice parameter of $c$=3.787 Å). Besides this, a dominant peak (2θ=38.62°) and two small intensity peaks are also detected (2θ = 40.87° and 2θ = 45.74°, see stars in Fig. 3(a)). Dominant peak corresponds to an out-of-plane parameter of $c_{perp}$=4.66(2) Å, much larger than the pseudocubic cell parameter of bulk BFO and it is associated with hyper strained tetragonal phase. The position of the two small peaks suggests the presence of two additional phases with out-of-plane parameters in the order of $c_{perp1}$=4.41(6) Å and $c_{perp2}$=3.96(7) Å, which are ascribed to the intermediate monoclinic structure and the rhombohedral phase respectively. While the intermediate monoclinic phase is associated to the formation of nanodomains [9, 12] the rhombohedral phase is the residue of the bulk counterpart phase [3]. Thickness value of the BFO/LAO film (t~45 nm) is determined from XRR curve (inset of Fig.3(a)). No impurity phases are detected, in agreement with the weak ferromagnetic moment measured at room-temperature (Fig.3(b)).

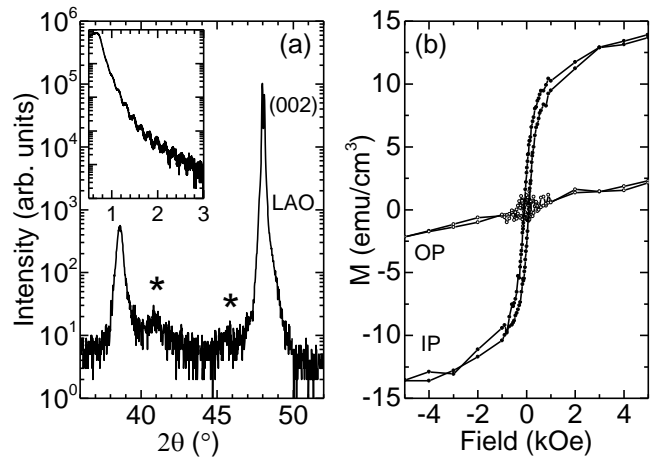

Fig. 3. (a) θ-2θ XRD scans of the (0 0 2) reflections for BFO/LAO thin film. In the inset, corresponding XRR data with thickness values of 45 nm. (b) Magnetization versus field corrected with the diamagnetic slope for the in-plane (IP) and out-of-plane field (OP) at room temperature.

Raman spectra of BFO film with ordered connected pits morphology are shown in Fig.4. Due to the fact that the film thickness is much smaller than the penetration depth of the beam, significant contribution from the LAO substrate is clearly observable in the BFO/LAO Raman spectra (Fig.4 (a)). For the purpose of probing the crystal structure of the BFO film, Raman spectra were collected for different sample orientation (as presented in Fig.4) in two polarization configurations.

Detailed analysis of the Raman intensity angular dependence [21] taking into account the twinning effects, indicates that Raman features observed at about 227, 273, 417 and 587 cm$^{-1}$ correspond to the *A'* symmetry modes, whereas peak at 368 cm$^{-1}$ correspond to *A''* symmetry mode of the monoclinic *Cc* crystal structure. These findings are in accordance with the density-functional calculations predicting that BFO structure under strain values higher than 4% become tetragonal-like with larger *c/a* ratio [22]. According to the first principles calculations, this structure has *Cc* symmetry with the base centered unit cell containing four formula) units, for which the factor group analysis predicts the existence of 13*A'* + 14*A''* Raman-active phonon modes [22]. Note that, a large number of modes ascribed to the *Cc* structure have been experimentally observed in the low-temperature Raman spectra of BiFeO$_3$ films commensurately grown on LaAlO$_3$ substrates and subjected to ~4.4% compressive strain, wherein it was indicated that 13 most intense modes (including those at 237, 282, 415, and 605 cm$^{-1}$) could be referred to the *A'* symmetry [23]. However, Himcinschi et al. [23] suggested that the modes at 225, 263 and 367 cm$^{-1}$ in the Raman spectrum of highly strained BiFeO$_3$ epitaxial films deposited on LaAlO$_3$ may correspond to *A''* symmetry mode of the *Cc* monoclinic structure [23]. Therefore, there is no doubt that BFO/LAO film whose Raman spectra are shown in Fig. 4 has *Cc* structure, but small differences in the positions of the identified modes in comparison with those referred in the literature [23] indicate that the structure of the film investigated here is monoclinically distorted in a specific way.

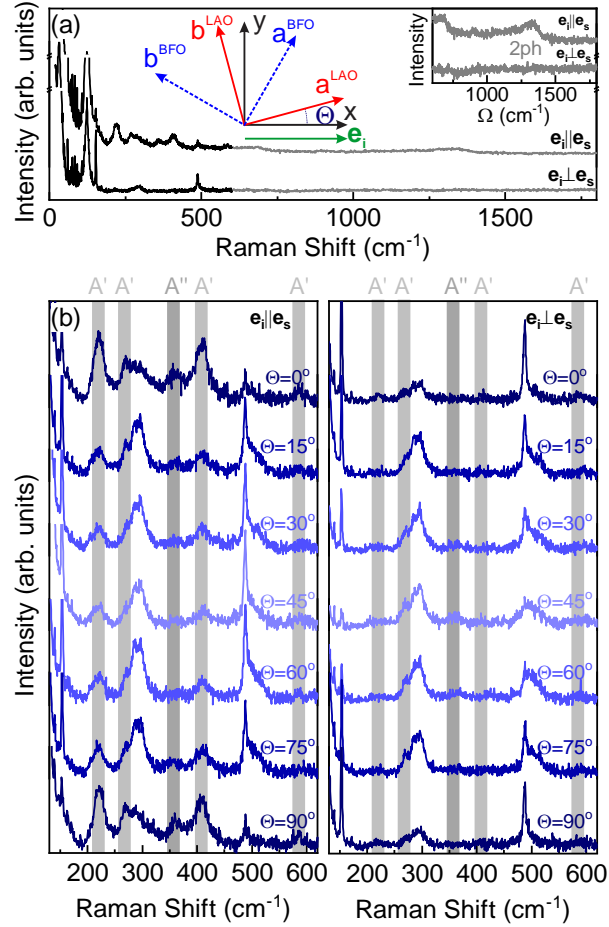

Fig. 4. (a) Raman spectra measured in parallel and cross polarization configurations for Θ=0. xy is the laboratory system at room temperature. Black and grey lines represent spectra obtained with binning 1 and 2 respectively. Inset: High energy region of the BFO/LAO Raman spectra. (b) BFO/LAO Raman spectra measured in parallel and cross polarization configurations for various sample orientations.

More details of the formation of the mixed-phase structures in the middle of the regular network line can be obtained from AFM topography shown in Fig.(5). The thin strip-line patterns could be visualized in Fig. 5(a), known in literature as asymmetric "saw-tooth" surface structure and identified as intimated mixture between highly distorted monoclinic phase (monoclinic version of highly tetragonal phase with $c/a$=1.23), detected by Raman scattering in Fig.(4) and an intermediate monoclinic phase ($c/a$=1.17), detected as small peak by XRD in Fig.(3(a)). The profile line of these strip-like patterns can be found in Fig.5(c)-(e). Typical strip-like patterns, disoriented around ~1° from the in-plane (100) and (010) axes, could be seen in green and blue squares in Fig.5(b). Corresponding AFM profile line indicate that they are tilted away from the surface normal for about 2.5-3° (intermediate monoclinic phase) and 1.5-2° (highly distorted monoclinic phase), in agreement with previous reports [9, 12]. The typical height difference between different phases (around 3 nm) [9, 24] is not always observed, as strongly perturbed with formation of regular grooved (for more details see Fig. S1 in supplementary material). In addition, the stripe-like formation could be observed also in our case between regular line-network and disoriented around ~150° from the in-plane (100) axes (red circles in Fig.5(b)) with very similar profile line (Fig.5(e)).

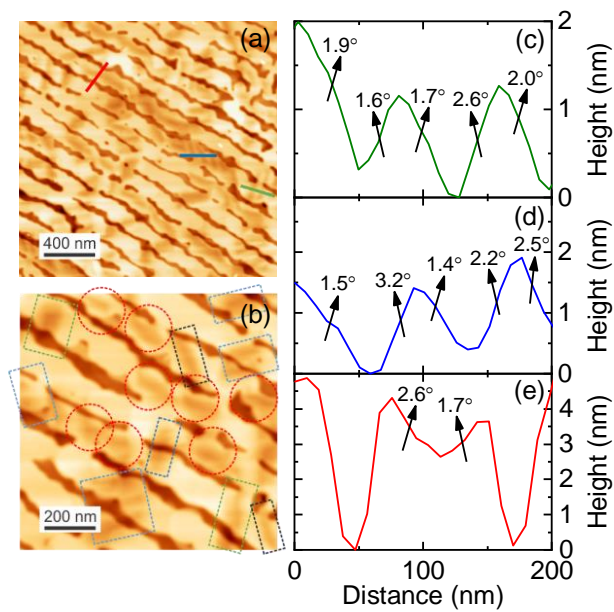

Fig. 5. The formation of nanodomains in the presence of regular line network. (a) AFM topography (2x2 µm$^2$) with profile lines. (b) Small area AFM topography (1x1 µm$^2$) with different orientation of detected nanodomains (expected orientation in blue and green rectangles and new nanodomains induced by line network in red circles). (c)-(e) Profile lines of nanodomains from (a).

Finally, the formation of the mixed-phase structures in the middle of the regular network is followed for different network orientations respected to [100] direction. In Fig. 6, AFM topography with corresponding Fast Fourier Transform is shown for BFO thin films grown on LAO with similar terrace width but with different orientation respected to substrate edges (6°<β<150°).

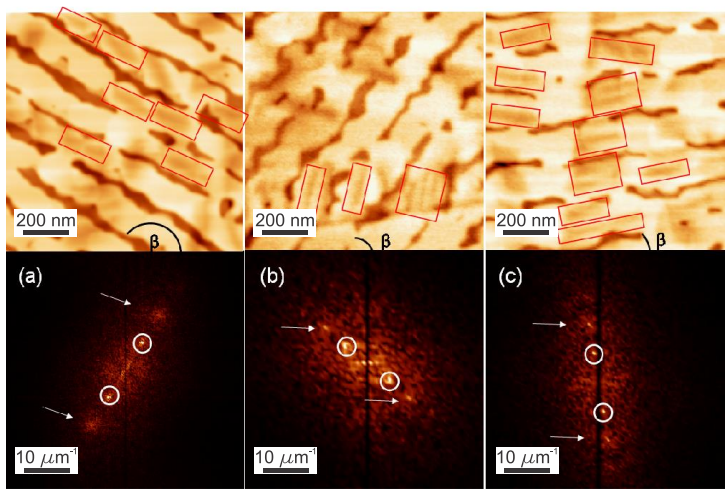

Fig. 6. AFM of BFO films grown of the top of LAO with similar miscut angle (0.12°<α<0.16°) with different orientation of nanostructured lines respected to [100] direction (a) β~150 °, (b) 52 ° and (c) 6 °. The corresponding FFT images are given below.

FFT of topographic images indicate the long-range order coming from of formation of groove network (two high intensity dots in circles in all cases). In addition, as mentioned before, two additional points (see arrows) are visible at doubled distance in $k$ space, suggesting the presence of additional structures in-between the regular line network. In all cases, the presence of "saw-tooth" surface structures in between the grooves could be identified in the corresponding AFM topography (see red rectangles and Fig S1 in supplementary). Noticeably, in the case of network line slightly disoriented from the in-plane (100) axes (β ~6°), the nanodomains form more easily in-between the grooves (see Fig 6(c)) as it is closer to their spontaneous configuration [9, 12, 24].

4. Conclusion

In summary, a regular nanostructured network is grown on the surface of BFO films, in close correlation with the step-terraces morphology of the underlying LAO substrate (miscut angle 0.12°<α<0.16°). Raman scattering detected the dominant contribution from the monoclinic version of highly strained tetragonal phase. XRD spectra also indicate the presence of additional polymorphs, rhombohedral and intermediate monoclinic phases. The stripe-like patterns, formed by intimate mixture between two monoclinic polymorphs, are visualized by AFM topography. Beside typical formation following (100) and (010) axes, some stripe-like patterns are detected also in-between the regular line network. Finally, the observed self-nanostructuration of the BFO surface opens a playground for tailoring structural polymorphs with strong potential interest for future applications in multiferroic nanomaterials.

Acknowledgements

This work has received funding from the European Union's Horizon 2020 research and innovation under the Marie Sklodowska-Curie grant agreement No. 645658 (DAFNEOX Project). A.P., V. F. and Z.K. thank Senzor-INFIZ (Serbia) for the cooperation provided during their respective secondments. E. P.-M. acknowledge financial support from the Spanish Ministry of Economy and Competitiveness through the "Severo Ochoa" Programme for Centres of Excellence in R&D (SEV-2015- 0496), and project MAT2015-71664 and SPINCURIOX (RTI2018-099960-B-I00). N. L., M. S., Z. K. and Z.V.P. acknowledge the support of the Serbian Ministry of Education, Science and Technological Development (Projects No III45018).

# Self-assembled line network in BiFeO$_3$ thin films

B. Colson[a], V. Fuentes [a], Z. Konstantinović[b], D. Colson[c], A. Forget[c], N. Lazarević[b], M. Šćepanović[b], Z. V. Popović[b], C. Frontera [a], Ll. Balcells [a], B. Martinez [a], A. Pomar[a]

[a] Institut de Ciencia de Materials de Barcelona, ICMAB-CSIC, Campus UAB, 08193 Bellaterra, Spain

[b] Center for Solid State Physics and New Materials, Institute of Physics Belgrade, Pregrevica 118, University of Belgrade, Serbia

[c] SPEC, CEA, CNRS UMR 3680, Université Paris-Saclay, 91191 Gif sur Yvette Cedex, France

**This file includes:**

Figure S1

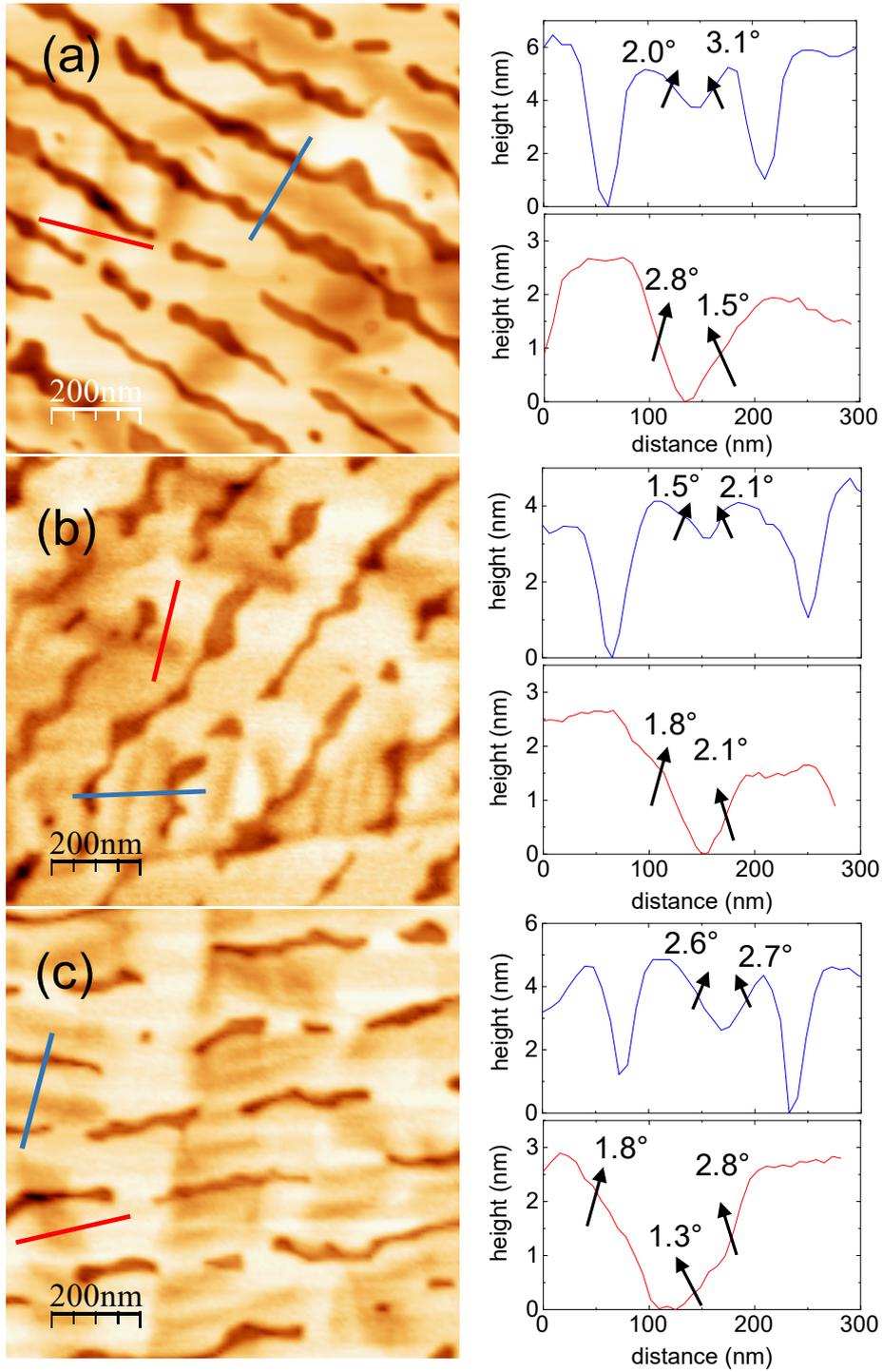

Fig. S1. AFM topology (1x1 μm²) and typical profile lines of BFO/LAO thin films (miscut angle in range of 0.12°<α<0.16°) with different orientation angle of nanostructured lines respected to [100] direction (a) β~150 °, (b) 52 ° and (c) 6 °.